\documentclass{aa}  

\usepackage{graphicx}
\usepackage{amsmath,amsfonts,amssymb}
\usepackage{natbib}

\usepackage{txfonts}
\usepackage{xcolor}

\usepackage{blindtext}
\usepackage{float}
\usepackage{dblfloatfix}
\usepackage{afterpage}
\usepackage{ifthen}
\usepackage[morefloats=12]{morefloats}

\usepackage{placeins}
\usepackage{multicol}
\usepackage[export]{adjustbox}\usepackage[breaklinks,colorlinks,citecolor=blue]{hyperref}
\bibpunct{(}{)}{;}{a}{}{,}
\usepackage[switch]{lineno}
\definecolor{linkcolor}{rgb}{0.6,0,0}
\definecolor{citecolor}{rgb}{0,0,0.75}
\definecolor{urlcolor}{rgb}{0.12,0.46,0.7}
\hypersetup{linktocpage}
\usepackage{bold-extra}
\usepackage{tabularx, booktabs}
\usepackage{amsmath}

\def\setsymbol#1#2{\expandafter\def\csname #1\endcsname{#2}}
\def\getsymbol#1{\csname #1\endcsname}

\def\Planck{\textit{Planck}}

\newbox\tablebox    \newdimen\tablewidth
\def\leaderfil{\leaders\hbox to 5pt{\hss.\hss}\hfil}
\def\tablenote#1 #2\par{\begingroup \parindent=0.8em
    \abovedisplayshortskip=0pt\belowdisplayshortskip=0pt
    \noindent
    $$\hss\vbox{\hsize\tablewidth \hangindent=\parindent \hangafter=1 \noindent
    \hbox to \parindent{$^#1$\hss}\strut#2\strut\par}\hss$$
    \endgroup}

\def\L2{\ifmmode L_2\else $L_2$\fi}

\def\DeltaT{\ifmmode \Delta T\else $\Delta T$\fi}
\def\deltat{\ifmmode \Delta t\else $\Delta t$\fi}
\def\fknee{\ifmmode f_{\rm knee}\else $f_{\rm knee}$\fi}
\def\Fmax{\ifmmode F_{\rm max}\else $F_{\rm max}$\fi}
\def\solar{\ifmmode{\rm M}_{\mathord\odot}\else${\rm M}_{\mathord\odot}$\fi}
\def\Msolar{\ifmmode{\rm M}_{\mathord\odot}\else${\rm M}_{\mathord\odot}$\fi}
\def\Lsolar{\ifmmode{\rm L}_{\mathord\odot}\else${\rm L}_{\mathord\odot}$\fi}
\def\inv{\ifmmode^{-1}\else$^{-1}$\fi}
\def\mo{\ifmmode^{-1}\else$^{-1}$\fi}
\def\sup#1{\ifmmode ^{\rm #1}\else $^{\rm #1}$\fi}
\def\expo#1{\ifmmode \times 10^{#1}\else $\times 10^{#1}$\fi}
\def\,{\thinspace}
\def\lsim{\mathrel{\raise .4ex\hbox{\rlap{$<$}\lower 1.2ex\hbox{$\sim$}}}}
\def\gsim{\mathrel{\raise .4ex\hbox{\rlap{$>$}\lower 1.2ex\hbox{$\sim$}}}}

\def\simprop{\mathrel{\raise .4ex\hbox{\rlap{$\propto$}\lower 1.2ex\hbox{$\sim$}}}}
\def\deg{\ifmmode^\circ\else$^\circ$\fi}
\def\pdeg{\ifmmode $\setbox0=\hbox{$^{\circ}$}\rlap{\hskip.11\wd0 .}$^{\circ}
          \else \setbox0=\hbox{$^{\circ}$}\rlap{\hskip.11\wd0 .}$^{\circ}$\fi}
\def\arcs{\ifmmode {^{\scriptstyle\prime\prime}}
          \else $^{\scriptstyle\prime\prime}$\fi}
\def\arcm{\ifmmode {^{\scriptstyle\prime}}
          \else $^{\scriptstyle\prime}$\fi}
\newdimen\sa  \newdimen\sb
\def\parcs{\sa=.07em \sb=.03em
     \ifmmode \hbox{\rlap{.}}^{\scriptstyle\prime\kern -\sb\prime}\hbox{\kern -\sa}
     \else \rlap{.}$^{\scriptstyle\prime\kern -\sb\prime}$\kern -\sa\fi}
\def\parcm{\sa=.08em \sb=.03em
     \ifmmode \hbox{\rlap{.}\kern\sa}^{\scriptstyle\prime}\hbox{\kern-\sb}
     \else \rlap{.}\kern\sa$^{\scriptstyle\prime}$\kern-\sb\fi}
\def\ra[#1 #2 #3.#4]{#1\sup{h}#2\sup{m}#3\sup{s}\llap.#4}
\def\dec[#1 #2 #3.#4]{#1\deg#2\arcm#3\arcs\llap.#4}
\def\deco[#1 #2 #3]{#1\deg#2\arcm#3\arcs}
\def\rra[#1 #2]{#1\sup{h}#2\sup{m}}

\def\dots{\relax\ifmmode \ldots\else $\ldots$\fi}
\def\WHzsr{\ifmmode $W\,Hz\mo\,sr\mo$\else W\,Hz\mo\,sr\mo\fi}
\def\mHz{\ifmmode $\,mHz$\else \,mHz\fi}
\def\GHz{\ifmmode $\,GHz$\else \,GHz\fi}
\def\mKs{\ifmmode $\,mK\,s$^{1/2}\else \,mK\,s$^{1/2}$\fi}
\def\muKs{\ifmmode \,\mu$K\,s$^{1/2}\else \,$\mu$K\,s$^{1/2}$\fi}
\def\muKRJs{\ifmmode \,\mu$K$_{\rm RJ}$\,s$^{1/2}\else \,$\mu$K$_{\rm RJ}$\,s$^{1/2}$\fi}
\def\muKHz{\ifmmode \,\mu$K\,Hz$^{-1/2}\else \,$\mu$K\,Hz$^{-1/2}$\fi}
\def\MJysr{\ifmmode \,$MJy\,sr\mo$\else \,MJy\,sr\mo\fi}
\def\MJysrmK{\ifmmode \,$MJy\,sr\mo$\,mK$_{\rm CMB}\mo\else \,MJy\,sr\mo\,mK$_{\rm CMB}\mo$\fi}
\def\microns{\ifmmode \,\mu$m$\else \,$\mu$m\fi}

\def\muK{\ifmmode \,\mu$K$\else \,$\mu$\hbox{K}\fi}
\def\microK{\ifmmode \,\mu$K$\else \,$\mu$\hbox{K}\fi}
\def\muW{\ifmmode \,\mu$W$\else \,$\mu$\hbox{W}\fi}
\def\kms{\ifmmode $\,km\,s$^{-1}\else \,km\,s$^{-1}$\fi}
\def\kmsMpc{\ifmmode $\,\kms\,Mpc\mo$\else \,\kms\,Mpc\mo\fi}
\providecommand{\sorthelp}[1]{}

\def\commanderthree{\texttt{Commander3}}

\newcommand{\A}[0]{\tens{A}}

\renewcommand{\L}[0]{\tens{L}}

\newcommand{\N}[0]{\tens{N}}

\newcommand{\BP}{\textsc{BeyondPlanck}}
\newcommand{\bp}{\textsc{BeyondPlanck}}
\newcommand{\cosmoglobe}{\textsc{Cosmoglobe}}
\newcommand{\Cosmoglobe}{\textsc{Cosmoglobe}}

\def\Cosmoglobe{\textsc{Cosmoglobe}}
\def\Planck{\textit{Planck}}
\def\WMAP{\textit{WMAP}}

\newcolumntype{C}{>{\centering\arraybackslash}m{0.3\textwidth}}

\newcolumntype{B}{>{\centering\arraybackslash}m{0.03\textwidth}}

\begin{document}

   \title{$N+2$ mapmaking for polarized CMB experiments}

   \institute{\small
Institute of Theoretical Astrophysics, University of Oslo, Blindern, Oslo, Norway\label{uio}
\and
Dipartimento di Fisica, Universit\`{a} degli Studi di Milano, Via Celoria, 16, Milano, Italy\label{milano}
\and
Università di Trento, Università degli Studi di Milano, CUP E66E23000110001\label{milano2}
\and
INFN sezione di Milano, 20133 Milano, Italy\label{milano3}
}

\author{\small
M.~Galloway\inst{\ref{uio}}\thanks{Corresponding author: M.~Galloway; \url{mathew.galloway@astro.uio.no}}
\and
H.~K.~Eriksen\inst{\ref{uio}}
\and
%E.~Gjerl\o w\inst{\ref{uio}}
%\and
%S.~K.~Næss\inst{\ref{uio}}
%\and
R.~M.~Sullivan\inst{\ref{uio}}
\and
D.~J.~Watts\inst{\ref{uio}}
\and
I.~K.~Wehus\inst{\ref{uio}}
\and
L.~Zapelli\inst{\ref{milano},\ref{milano2},\ref{milano3}}
}

   %\institute{Institute of Theoretical Astrophysics, University of Oslo, Blindern, Oslo, Norway}
  
   % Shortened title, author list for top of page 
   \titlerunning{$N+2$ mapmaking}
   \authorrunning{Galloway et al.}

   \date{\today} 
   
   \abstract{We introduce $N+2$ mapmaking as a novel approach to constructing maps in both intensity and polarization for multi-detector CMB data. The motivation behind this method is two-fold: Firstly, it provides individual temperature detector maps from a multi-detector set, which may be useful for component separation purposes, in particular for line emission reconstruction. Secondly, it simultaneously outputs coadded polarization maps with minimal temperature-to-polarization leakage sensitivity. Algorithmically speaking, the $N+2$ mapmaker is closely related to the `spurious mapmaking' algorithm pioneered by the \WMAP\ team, but rather than solving for a spurious $S$ map together with the three normal Stokes $IQU$ parameters, we solve for $N$ temperature maps and two Stokes ($Q$ and $U$) parameters per pixel. The result is a statistically coherent set of physically meaningful per-detector temperature maps, each with slightly different bandpasses as defined by each detector, combined with coadded polarization maps. We test this approach on \Planck\ Low Frequency Instrument (LFI) 30 GHz data, and  find that the \Planck\ scanning strategy is too poorly cross-linked to allow for a clean separation between temperature and polarization. However, noting that pairs of detectors within a single horn are strongly anti-correlated, we anticipate that solving for horn maps, as opposed to individual detector maps, may provide an optimal compromise between noise and temperature-to-polarization leakage minimization. When applied to simulated data with a rotating half-wave plate, for which the polarization angle coverage is greatly improved, the algorithm performs as expected. For such experiments, the $N+2$ mapmaker offers the option of constructing multi-detector maps with minimal temperature-to-polarization leakage, which will be useful for detecting and mitigating low-level systematic effects. We conclude that $N+2$ mapmaking is a useful tool for adjusting the granularity of the temperature map decomposition of a given experiment, while still producing joint high signal-to-noise ratio polarization maps with minimal temperature-to-polarization leakage.
   }
   
   \maketitle
%\setcounter{tocdepth}{2}
%\tableofcontents
   
% INTRODUCTION
%-------------------------------------------------------------------
\section{Introduction}

The problem of bandpass mismatch is important when combining multiple cosmic microwave background (CMB) detectors into a single polarization map \citep[e.g.,][]{page:2007,lfi2015,BP09}. Small differences in the bandpass of individual detectors can result in major disagreements about signal amplitudes in regions with bright foregrounds with non-thermal spectra, as each detector effectively observes a different sky signal. These disagreements then lead to temperature-to-polarization leakage as the differences in temperature are incorrectly interpreted as a polarized signal instead of as bandpass differences by the mapmaking algorithm. For frequency channels that exhibit bright line emission, such as the \Planck\ HFI 100, 217 and 353\,GHz channels, which all contain significant CO emission, this issue is particularly pressing \citep[e.g.,][]{hfi_processing:2013}. Indeed, for such channels it is  advantageous to have access to individual temperature maps for each detector, in order to maximize the sensitivity with respect to the CO signal \citep{planck_co:2014}, while at the same time co-adding the polarization components of the time-ordered data (TOD) into common polarization maps.

Multiple approaches have been proposed to mitigate these issues in the literature already. Starting with the problem of minimizing the temperature-to-polarization leakage, the most obvious solution is to bin the data for each detector independently into separate maps. However, this requires a sufficiently cross-linked scanning strategy, and this is usually only possible for experiments with a spinning half-wave plate \citep[e.g.,][]{abs:2016}. For poorly or moderately cross-linked scanning strategies, individual detector maps are not usually directly usable, but rather require combination in post-processing. A second common approach is to explicitly correct for the bandpass of each detector by assuming an explicit sky model that describes the spectral energy density of each component \citep[e.g.,][]{planck_fg:2015}, and subtract the predicted bandpass differences for each detector, either in terms of spatial templates \citep[e.g.,][]{lfi2015} or directly at the level of time-ordered data \citep{bp01}. A third approach is to solve for an additional so-called `spurious map' together with the regular Stokes parameters, by adding an additional component to the mapmaking vector that accounts for the excess signal seen by an individual detector \citep{spurious}, at the cost of an increased conditional number in the mapmaking equation, and thereby also somewhat higher noise. A fourth approach is to combine the latter two methods, by adjusting the bandpasses used in the explicit modelling approach by minimizing the amplitudes of the spurious maps \citep{BP09}. As for producing individual detector maps for component separation purposes, this is usually done as a separate post-processing step, in which instrumental parameters (e.g., gain and correlated noise) are fixed to their global values, and individual detector TOD are binned into separate maps, either with or without explicit depolarization \citep[e.g.,][]{npipe}.

In this paper, we introduce a new variation of the standard mapmaker that addresses both of these issues simultaneously. This method is closely inspired by the spurious mapmaking algorithm introduced by \citet{spurious}. However, rather than solving for a spurious map, $S$, together with common temperature and polarization maps, we solve for $N$ individual temperature maps, where $N$ is the number of detectors, in addition to common Stokes $Q$ and $U$ parameters. We call this $N+2$ mapmaking. In principle, this should allow for the exact bandpass to be used for each detector map, and thereby maximize the available information for component separation purposes, while at the same time the polarization data are combined into co-added maps with a maximal signal-to-noise ratio and minimal temperature-to-polarization leakage. 

We present the algorithm in Sect.~\ref{sec:mapmaking}. In Sect.~\ref{sec:lfi}, we start by naively testing the algorithm on real \Planck\ 30\,GHz data, and find unfortunately, as already noted by numerous authors \citep[e.g.,][]{planckScan}, that the \Planck\ scanning strategy is not sufficiently cross-linked to support a robust decomposition into individual temperature maps. However, we also note that detectors in a single horn are strongly anti-correlated, and this suggests that combining detectors pairwise into horn maps may produce stable results; indeed, this decomposition has already been demonstrated to work well by \citet{npipe}. Next, we show that the $N+2$ algorithm works well on a simulated data set that corresponds to the \Planck\ scan path but augmented with a spinning half-wave plate. In Sect.~\ref{sec:bandpass}, we illustrate the advantages of the algorithm in terms of temperature detector difference maps and temperature-to-polarization suppression efficiency. We conclude in Sect.~\ref{sec:conclusions}, and offer some avenues to exploit this technique in the future. Finally, we note that the original motivation for this work was to prepare for a future end-to-end Bayesian analysis of the \Planck\ HFI data, similar to those already performed for LFI \citep{bp01}, \WMAP\ \citep{watts2023_dr1}, and DIRBE \citep{CG02_01} by the \bp\ and \Cosmoglobe\footnote{\url{http://cosmoglobe.uio.no}} collaborations, and this paper represents the first product from the OpenHFI collaboration.

\begin{figure*}[p]
  \centering
  \includegraphics[width=0.48\textwidth]{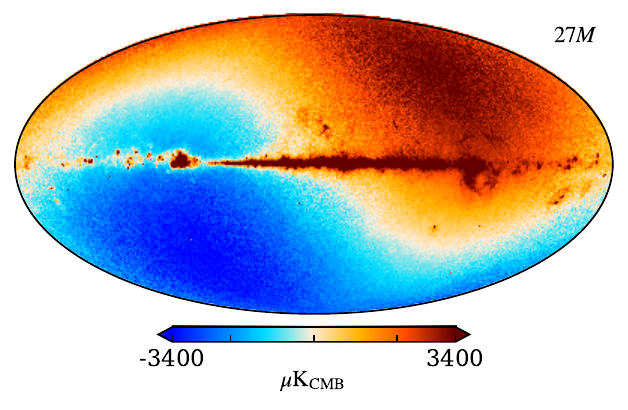}
  \includegraphics[width=0.48\textwidth]{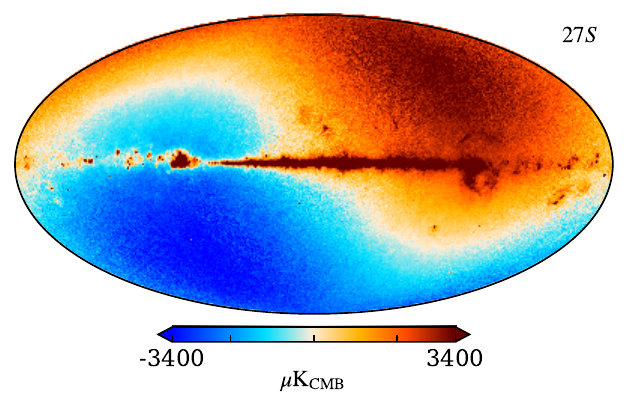}\\
  \includegraphics[width=0.48\textwidth]{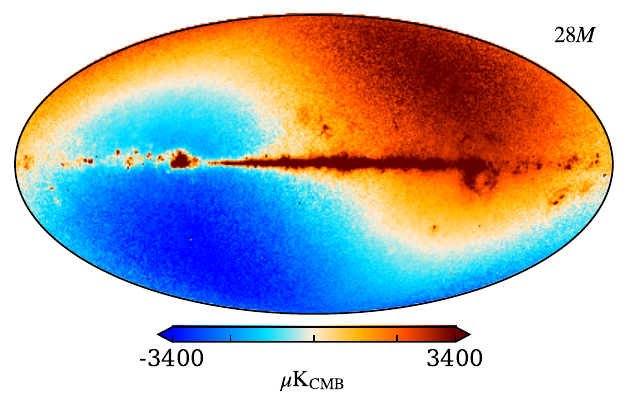}
  \includegraphics[width=0.48\textwidth]{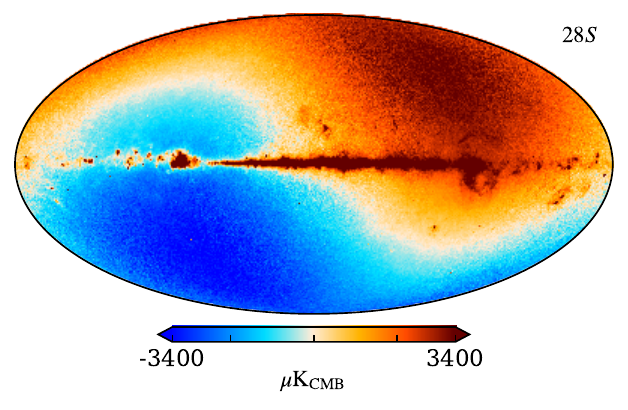}\\
  \includegraphics[width=0.48\textwidth]{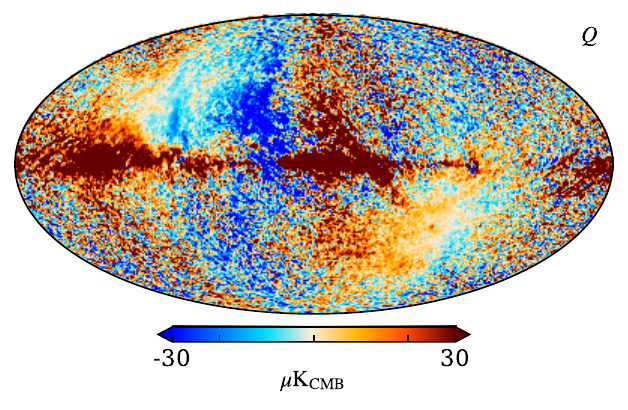}
  \includegraphics[width=0.48\textwidth]{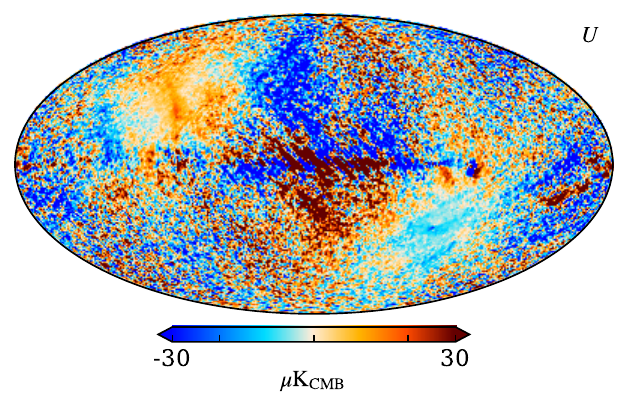}\\
  \includegraphics[width=0.48\textwidth]{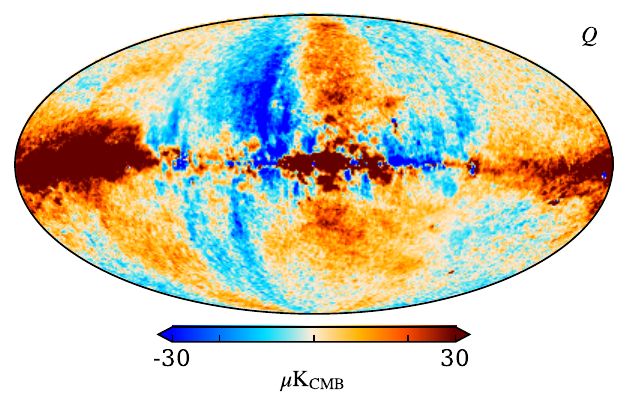}
  \includegraphics[width=0.48\textwidth]{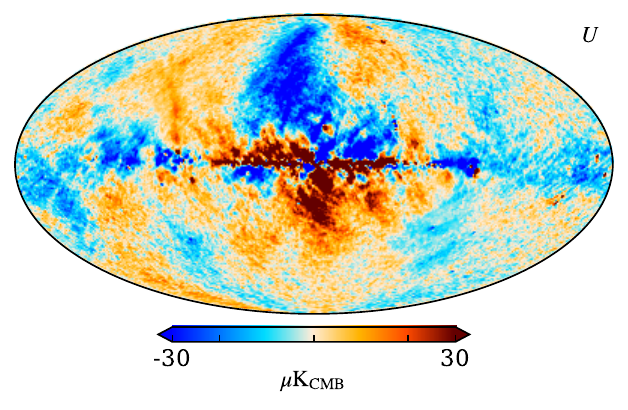}\\
  \caption{Temperature and polarization maps for the LFI 30\,GHz data produced using $N+2$ mapmaking. The top four panels shows temperature maps for individual detectors, and the third row shows the combined N+2 $Q$ and $U$ maps. The bottom row shows corresponding $Q$ and $U$ maps produced from a traditional binned mapmaker that co-adds all data into a common temperature map.}
  \label{fig:maps}
\end{figure*}

\section{Mathematical Description}
\label{sec:mapmaking}

The mapmaking problem is often expressed in the literature as \citep[e.g.,][]{de_Gasperis_2005}
\begin{equation}
d_t = \A_{t,p}s_p + n_t,
\end{equation}
where $t$ and $p$ denote time sample and pixel, respectively; $d_t$ is the detector timestream; $\A$ is the pointing matrix; $s$ is the signal vector; and $n$ represents zero-mean instrumental Gaussian noise with covariance $\N$. In the case of $N+2$ mapmaking, we write $d_t$ as 
\begin{equation}
d_t = \begin{pmatrix}
d_t^1\\ d_t^2\\ \vdots \\ d_t^i\\
\end{pmatrix},
\end{equation}
where $i$ indexes detectors, while the sky signal for a single pixel, $s_p$, is written as
\begin{equation}
s_p = \begin{pmatrix}
I_{1,p}\\
I_{2,p}\\
\vdots\\
I_{i,p}\\
Q_p\\
U_p\\
\end{pmatrix}.
\end{equation}
This differs from the standard approach by allowing individual temperature maps per detector. Our data model for the timestream of a single detector $i$ then becomes
\begin{equation}
d_{i,p,t} = I_{i,p} + Q_p \cos2\phi_i(t) + U_p \sin2\phi_i(t),
\label{eq:datamodel}
\end{equation}
where $\phi$ is the polarization angle of the detector. In this model, the $Q$ and $U$ terms are common between all detectors, and thus independent of $i$.

Based on this model, we generalize the pointing matrix such that it maps the correct detector to the correct temperature map,
\begin{equation}
\A_{t,p} = \begin{pmatrix}
1 & 0 & \cdots & 0 & \cos2\phi_1 & \sin2\phi_1 \\
0 & 1 & \cdots & 0 & \cos2\phi_2 & \sin2\phi_2 \\
\vdots & \vdots & \ddots & \vdots & \vdots & \vdots \\
0 & 0 & \cdots & 1 & \cos2\phi_i & \sin2\phi_i \\
\end{pmatrix},
\end{equation}
where, for notational ease, we have dropped the explicit function of time for the polarization angles. The standard Generalized Least Squares (GLS) solution to the mapmaking equation is as usual given by
\begin{equation}
\tilde{s}_p = (\A^T \N^{-1}\A)^{-1} \A^T\N^{-1}d.
\label{eq:gls}
\end{equation}

During the mapmaking process, we must accumulate the quantity $\A^T\N^{-1}\A$. Assuming $\N$ to be diagonal and given by
\begin{equation}
\N = \begin{pmatrix}
\sigma^2_1 & 0 & \cdots & 0 \\
0 & \sigma^2_2 & \cdots & 0 \\
\vdots & \vdots & \ddots & \vdots \\
0 & 0 & \cdots & \sigma^2_i \\
\end{pmatrix},
\end{equation}
this quantity expands into
\begin{equation}
\begin{tiny}
\sum_t
\begin{pmatrix}
\frac{1}{\sigma_1^2} & 0 & \cdots &
\frac{\cos2\phi_1}{\sigma_1^2} & \frac{\sin2\phi_1}{\sigma_1^2} \\

0 & \frac{1}{\sigma_2^2} & \cdots &
\frac{\cos2\phi_2}{\sigma_2^2} & \frac{\sin2\phi_2}{\sigma_2^2} \\

\vdots & \vdots & \ddots & \vdots & \vdots \\

\frac{\cos2\phi_1 }{\sigma_1^2} & \frac{\cos2\phi_2}{\sigma_2^2} & \cdots & 
\sum_{i} \frac{\cos^2 2\phi_i}{\sigma_i^2} & \sum_{i} \frac{\sin2\phi_t \cos2\phi_i}{\sigma_i^2}  \\

\frac{\sin2\phi_1}{\sigma_1^2} & \frac{\sin2\phi_2}{\sigma_2^2} & \cdots &
\sum_{i} \frac{\sin2\phi_i \cos2\phi_i}{\sigma_i^2}  & \sum_{i} \frac{ \sin^2 2\phi_i}{\sigma_i^2}
\\
\end{pmatrix}
\end{tiny}
\end{equation}
for a given pixel, where the sum over $t$ represents the sum over all observation hitting a particular pixel $p$. Simultaneously, we can also accumulate the vector $\A^T\N^{-1}d$ which has the form
\begin{equation}
\sum_t
\begin{pmatrix}
\frac{d_{1}}{\sigma_1^2} \\
\frac{d_{2}}{\sigma_2^2}\\
\vdots\\
\frac{d_{i}}{\sigma_i^2}\\
\sum_i \frac{d_{i}\cos2\phi_i}{\sigma_i^2}\\
\sum_i \frac{d_{i}\sin2\phi_i}{\sigma_i^2}\\
\end{pmatrix}.
\end{equation}
Once these quantities are computed for all observations over the entire mission, it remains to simply invert the first matrix and front multiply the data vector to obtain the sky vector $\tilde{s}$, as described by Eq.~\ref{eq:gls}.

\begin{figure}[t]
  \centering
  \includegraphics[width=0.47\linewidth]{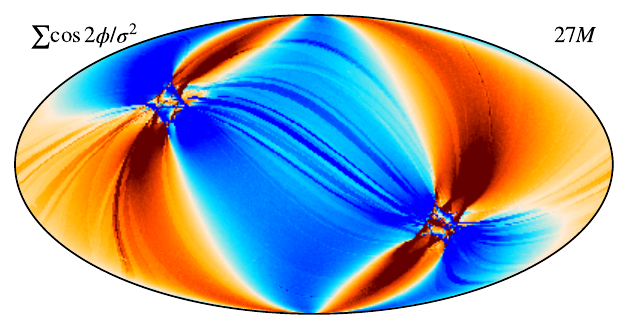}
  \includegraphics[width=0.47\linewidth]{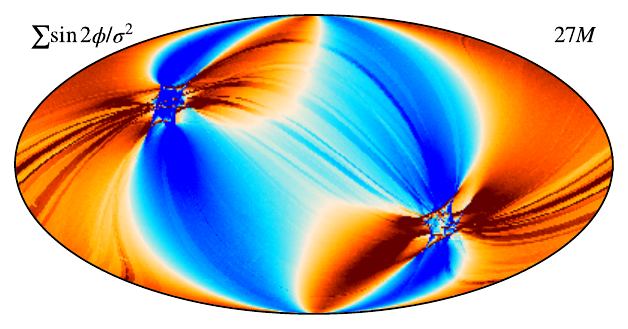}\\
  \includegraphics[width=0.47\linewidth]{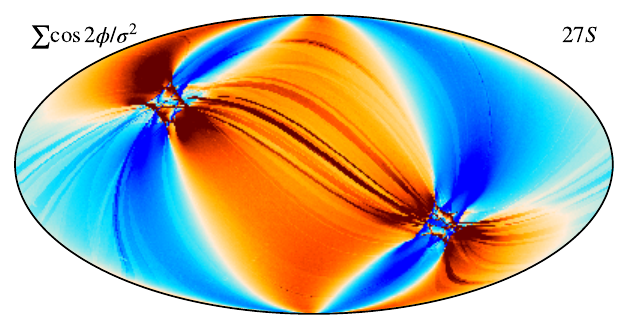}
  \includegraphics[width=0.47\linewidth]{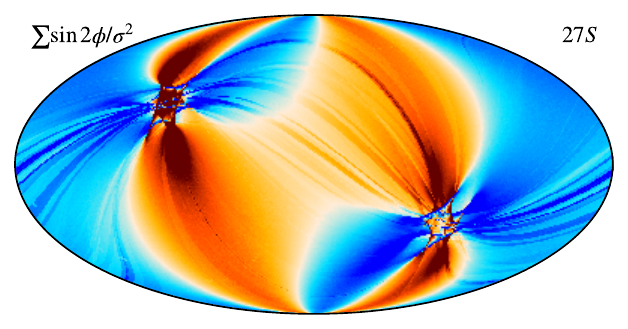}\\
  \includegraphics[width=0.47\linewidth]{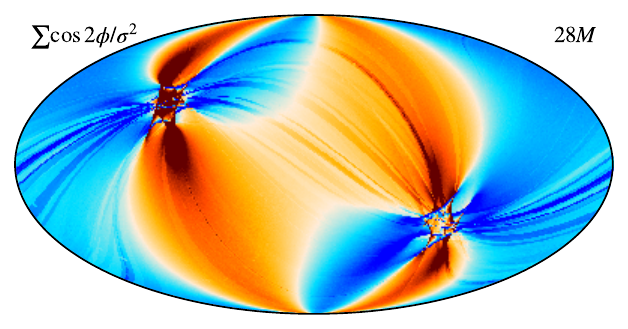}
  \includegraphics[width=0.47\linewidth]{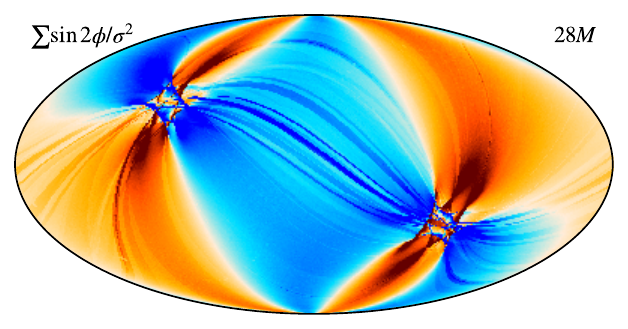}\\
  \includegraphics[width=0.47\linewidth]{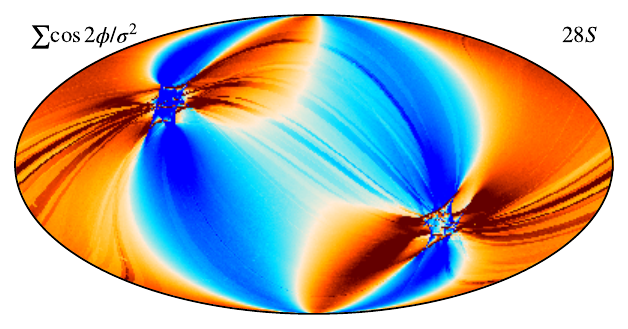}
  \includegraphics[width=0.47\linewidth]{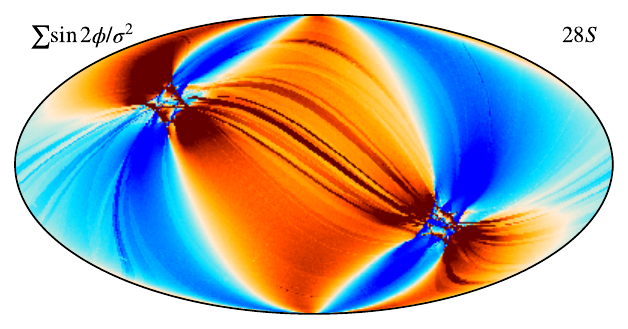}\\
  \includegraphics[width=0.47\linewidth]{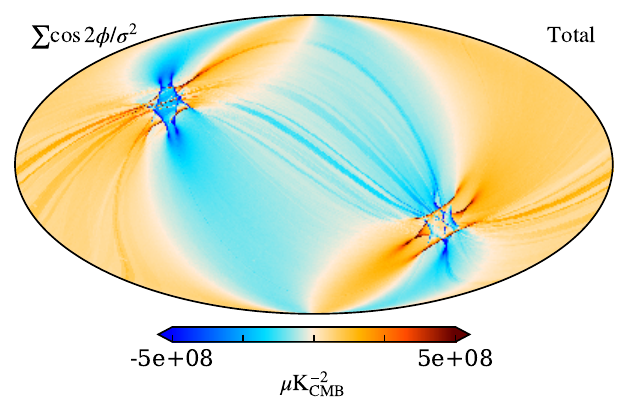}
  \includegraphics[width=0.47\linewidth]{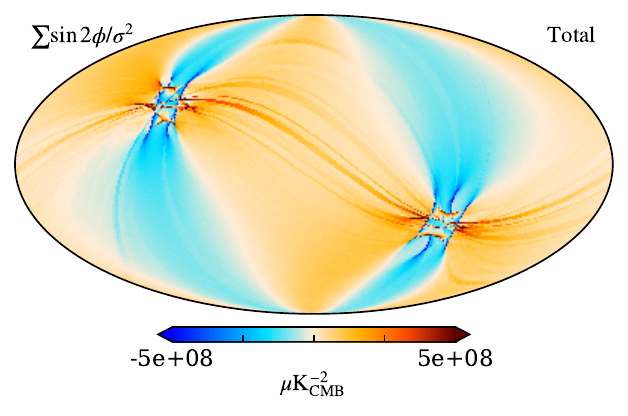}\\
  \caption{Polarization terms from the four 30\,GHz LFI detectors (rows 1--4) and the combined term used by the traditional binned mapmaker (bottom row). The left column shows the Stokes $Q$ term, $(\frac{1}{\sigma})^2 \cos2\phi$, and the right column shows the Stokes $U$ term, $(\frac{1}{\sigma})^2 \sin2\phi$.}
  \label{fig:polangles}
\end{figure}

\section{Application to \Planck\ LFI 30\,GHz}
\label{sec:lfi}

\begin{figure*}
  \centering
  \includegraphics[width=0.49\textwidth]{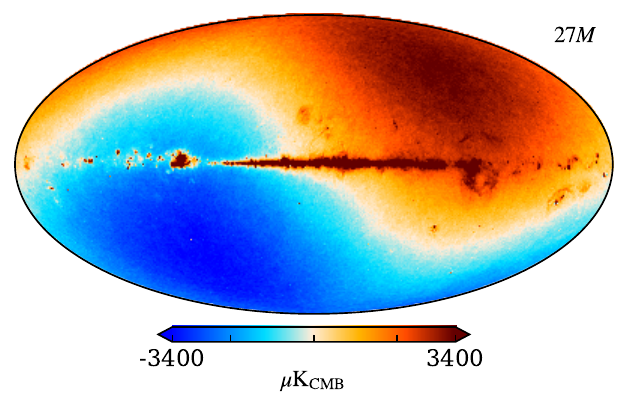}
  \includegraphics[width=0.49\textwidth]{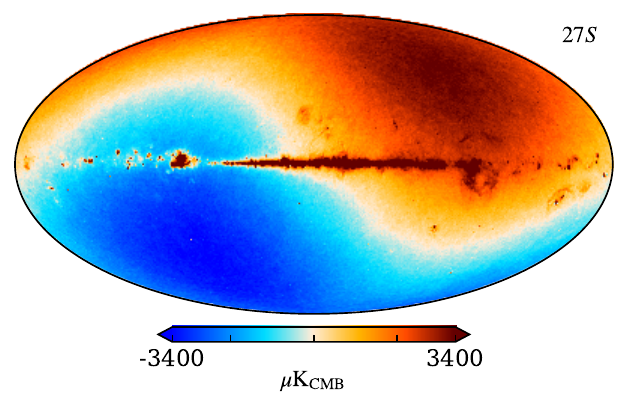}\\
  \includegraphics[width=0.49\textwidth]{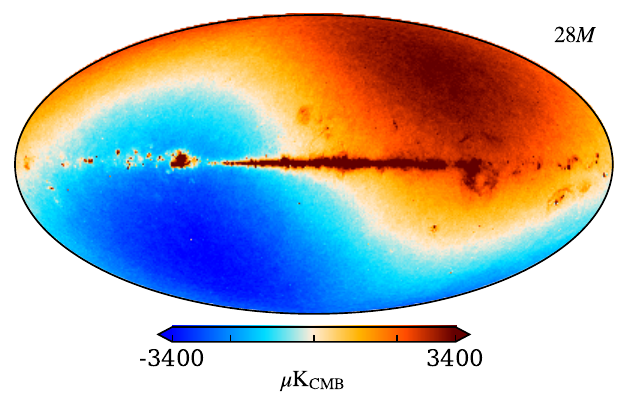}
  \includegraphics[width=0.49\textwidth]{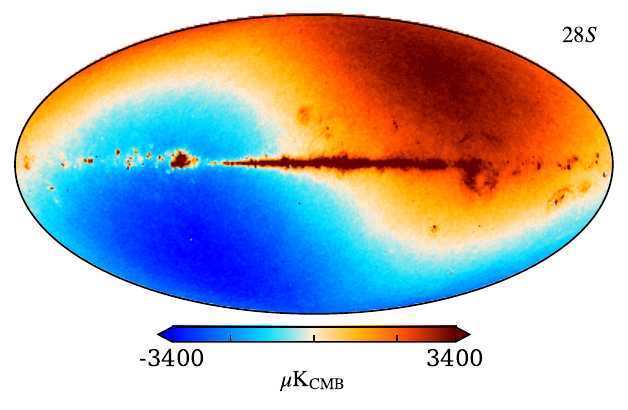}\\
  \includegraphics[width=0.49\textwidth]{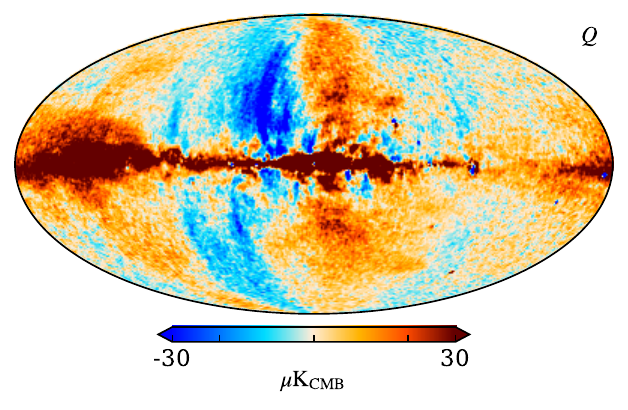}
  \includegraphics[width=0.49\textwidth]{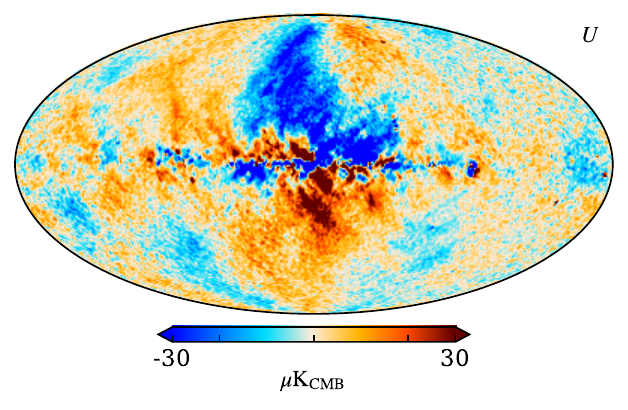}\\
  \caption{Simulated temperature and polarization maps of the LFI 30\,GHz data and an idealized spinning half wave plate, produced using $N+2$ mapmaking. The top four panels show the temperature maps, and the bottom row shows the combined $Q$ and $U$ maps.}
  \label{fig:sim}
\end{figure*}

\begin{figure*}
  \centering
  \includegraphics[width=0.49\textwidth]{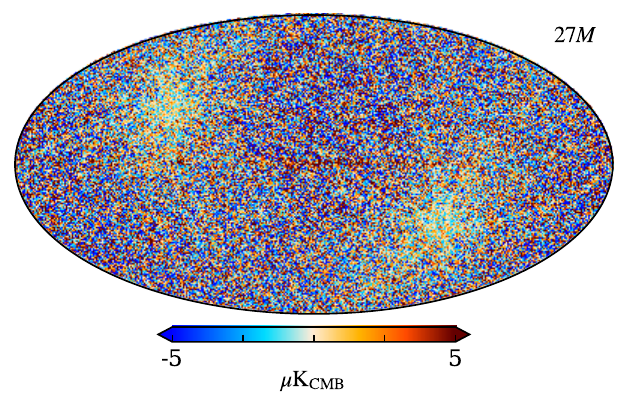}
  \includegraphics[width=0.49\textwidth]{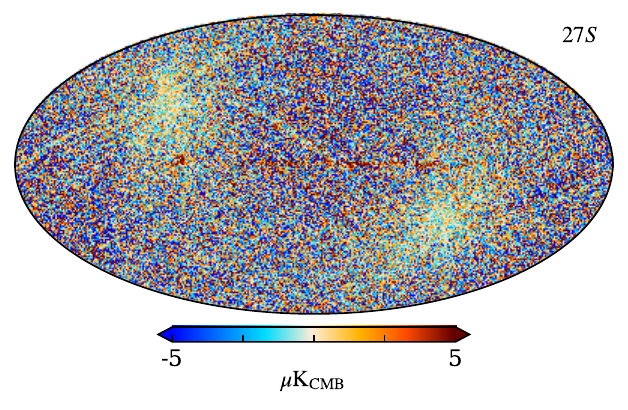}\\
  \includegraphics[width=0.49\textwidth]{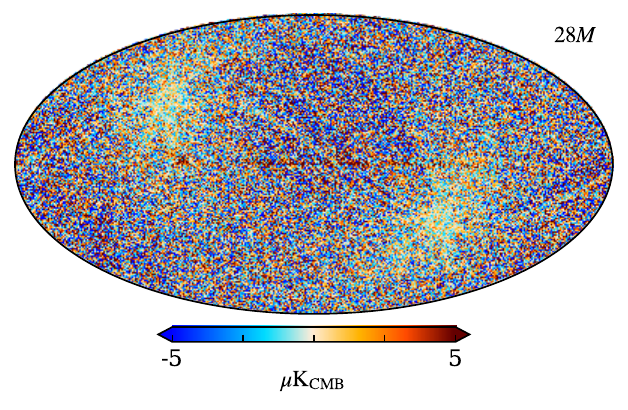}
  \includegraphics[width=0.49\textwidth]{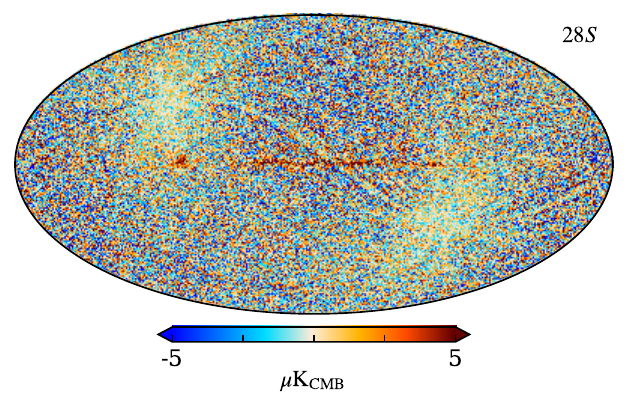}\\
  \includegraphics[width=0.49\textwidth]{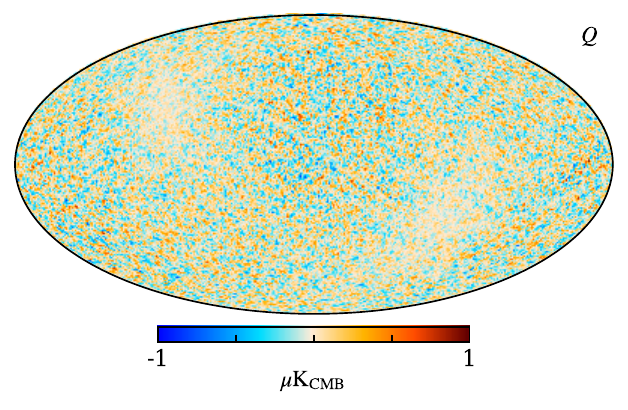}
  \includegraphics[width=0.49\textwidth]{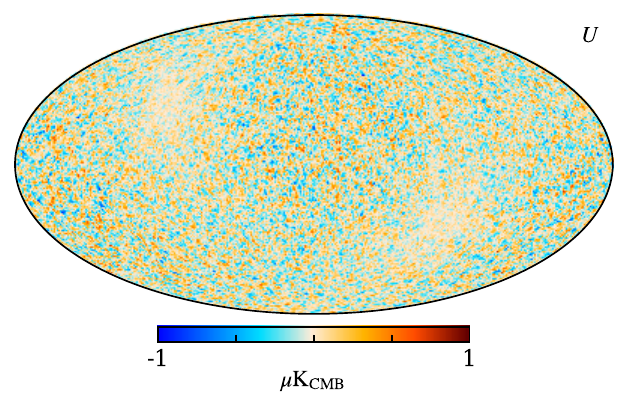}\\
  \caption{Input sky minus output maps for the simulated LFI 30\,GHz sky as generated using $N+2$ mapmaking in \commanderthree\ . The top four panels show the temperature maps, and the bottom row shows the combined $Q$ and $U$ maps.}
  \label{fig:sim_diff}
\end{figure*}

To understand the behaviour of the new $N+2$ mapmaking algorithm, we start by applying it to a well-known case, namely the \Planck\ LFI 30\,GHz data. In particular, we implement this algorithm within the Bayesian \commanderthree\ \citep{bp03} Gibbs sampling code developed by the \BP\ and \cosmoglobe\ collaborations \footnote{All code used in this analysis is publicly available at \url{https://github.com/Cosmoglobe/Commander}} \citep{bp01, watts2023_dr1}, and the raw TOD data are processed in an identical manner as in those works with respect to calibration, correlated noise, etc. Only the signal-plus-white noise residual TOD are fed to the $N+2$ mapmaker, and our new contribution thus replaces the full-frequency binned mapmaker presented by \citet{BP10}. 

To briefly summarize this general algorithmic approach, we start by defining a single parametric model for both the sky and instrument, and for LFI this is currently given by
\begin{equation}
  \begin{split}
    d_{j,t} = g_{j,t}&\tens{P}_{tp,j}\left[\tens{B}^{\mathrm{symm}}_{pp',j}\sum_{c}
      \tens{M}_{cj}(\beta_{p'}, \Delta_{\mathrm{bp}})a^c_{p'}  + \tens{B}^{\mathrm{asymm}}_{j,t}\left(\vec{s}^{\mathrm{orb}}_{j}  
      + \vec{s}^{\mathrm{fsl}}_{t}\right)\right] + \\
%    + s^{\mathrm{fsl}}_{j,t} + s^{\mathrm{mono}}_{j}\right] + \\
    + &n^{\mathrm{corr}}_{j,t} + n^{\mathrm{w}}_{j,t}.
  \end{split}
  \label{eq:todmodel}
\end{equation}
The details of this model are discussed in \citep{bp01}, but we note that the terms directly relevant to mapmaking are the pointing matrix $\tens{P}_{tp,j}$; the sky signal, which is the sum over sky components $\sum_{c} \tens{M}_{cj}(\beta_{p'}, \Delta_{\mathrm{bp}})a^c_{p'}$; and the white noise $n^{\mathrm{w}}_{j,t}$. The rest of the terms (like the sidelobes or correlated noise) are allowed by the Gibbs sampling algorithm \citep{gibbs} to be assumed fixed during mapmaking, and thus can be treated as deterministic contaminants and removed as a pre-processing step. This approach greatly simplifies the formal mapmaking process, as we do not need to mitigate correlated noise or other systematics during mapmaking, and can simply bin the TOD per pixel to average down the white noise, using the approach detailed in Sect.~\ref{sec:mapmaking}.

Applying this method to the four LFI 30\,GHz detectors (denoted 27M, 27S, 28M, and 28S), the $N+2$ mapmaker produces four distinct temperature maps, seen in Fig.~\ref{fig:maps}. In the first and second rows, we show the temperature maps, and the third row shows the combined $Q$ and $U$ maps. For comparison, the bottom row shows the corresponding $Q$ and $U$ maps produced from the traditional binned mapmaker, which have been extensively verified by multiple implementations of the algorithm over many years \citep[e.g.,][]{lfi2013,lfi2015,lfi2018}. Comparing the maps in the third and fourth rows in Fig.~\ref{fig:maps}, we see already at a purely visual level that the new $N+2$ mapmaker does not produce maps that are competitive with the traditional approach for the \Planck\ 30\,GHz data, neither in terms of overall noise levels nor temperature-to-polarization leakage. We investigate the origin of this problem in the next section.

\subsection{Polarization Angle Coverage}

The root cause of the polarization problems seen in Fig.~\ref{fig:maps} is the limited polarization angle coverage of the \Planck\ scanning strategy. When the \Planck\ mission was originally designed, it was optimized for thermal stability with respect to intensity reconstruction, which resulted a scan path following nearly great circles in Ecliptic coordinates, modulated by a slow cycloidal procession to improve coverage at the ecliptic poles \citep{planckScan}. While this scanning strategy did ensure that the \Planck\ instrument could meet its thermal requirements, it also had unfortunately consequences for polarization reconstruction. The main problem is simply that each detector sees each pixel on the sky with a very limited range of polarization angles, $\phi$. This, in turn, results in a poorly conditioned matrix $\A^T\N^{-1}\A$, as defined in Sect.~\ref{sec:mapmaking}. This is visualized in Fig.~\ref{fig:polangles}, where we plot the off-diagonal temperature-polarization cross terms, $(\frac{1}{\sigma_i})^2 \cos2\phi_i$ and $(\frac{1}{\sigma_i})^2 \sin2\phi_i$, for the four 30\,GHz detectors in the top four rows. For comparison, the bottom row shows the same quantity for the full co-added frequency channel map, which corresponds to the traditional binned mapmaker. 

For an experiment with a perfectly uniform polarization coverage, these maps are consistent with zero, as the sine and cosine terms cancel when averaged over the polarization angle $\phi$. As a result, the coupling matrix, $\A^T\N^{-1}\A$, becomes diagonal, and the overall condition number is defined by the noise level alone. For the \Planck\ scanning strategy, this is not the case. Rather, we immediately see from the bottom row in Fig.~\ref{fig:polangles} that the magnitude of the coadded coupling matrix is almost an order of magnitude smaller than any of the individual detectors. Even worse, we also note that pairs of detectors, such as 28M and 28S, have a very similar spatial morphology, but with opposite signs. As a result, the mean condition number of the per-detector coupling matrix is about 100, while it is about 2 for the co-added case. This means effectively that the white noise variance of the individual detector maps is boosted by a factor of 50, and this is what is seen visually in the bottom two rows of Fig.~\ref{fig:maps}. In practice, this means that the $N+2$ mapmaking algorithm is not immediately useful for \Planck\ in its most direct way. On the other hand, these calculations do also suggest that building horn maps (by coadding pairs of detectors within a single horn) may be a viable strategy for future analysis.

\subsection{Simulations with a spinning half-wave plate}
\label{sec:sim}

Next, we apply the $N+2$ mapmaker to a case with nearly perfect polarization angle coverage, by replacing the real LFI 30\,GHz data above with a \commanderthree-based TOD simulation, as described in \commanderthree\ \citep{BP04}. Intuitively speaking, this amounts to replacing the real data with a random realization drawn from the data model described by Eq.~\ref{eq:todmodel}. However, to mitigate the poor polarization coverage discussed above, we replace the polarization angle in each sample with a random orientation, which essentially corresponds to adding an ideal infinitely fast spinning half-wave plate to the \Planck\ optical system. The resulting maps are shown in Fig.~\ref{fig:sim}, and the corresponding differences with respect to the known input sky map is shown in Fig.~\ref{fig:sim_diff}. With this modification, we see a low-amplitude residual along the Galactic plane in temperature at the level of $\mathcal{O}(10^{-4})$; because there is a finite number of samples hitting each pixel, the off-diagonal coupling matrix is not perfectly zero, and a small level of bandpass-induced temperature-to-polarization leakage therefore still leaks through in the very brightest foreground emission parts of the sky. However, both the rest of the temperature maps and the polarization maps are all consistent with noise, and this demonstrates that the $N+2$ algorithm does work as intended.

\section{Bandpass reconstruction and leakage mitigation}
\label{sec:bandpass}

\begin{figure*}
\includegraphics[width=0.31\textwidth]{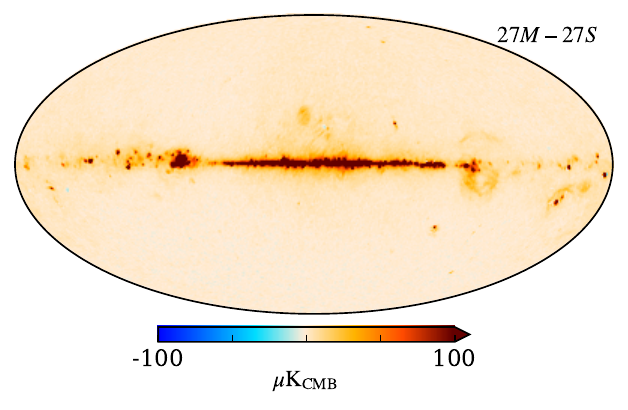} 
\includegraphics[width=0.31\textwidth]{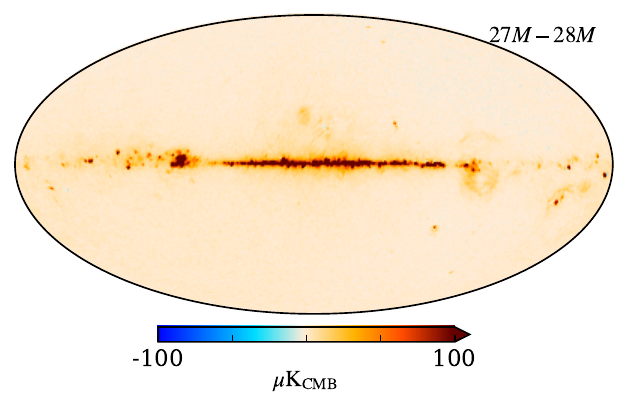} 
\includegraphics[width=0.31\textwidth]{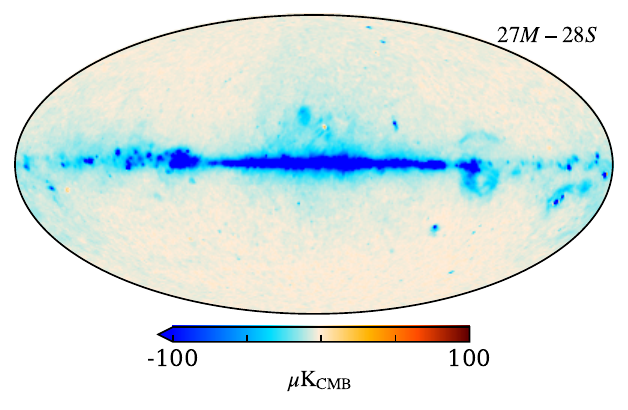}\\
\includegraphics[width=0.31\textwidth]{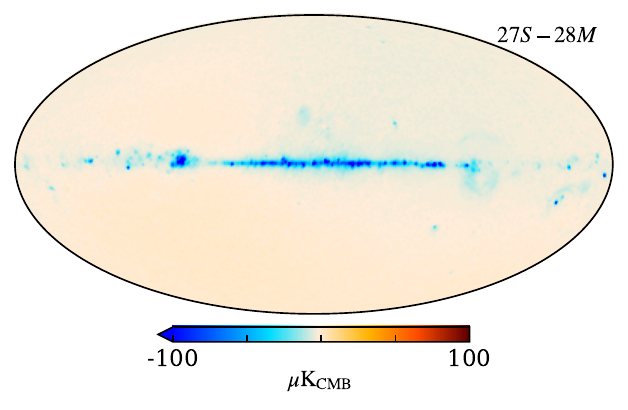} 
\includegraphics[width=0.31\textwidth]{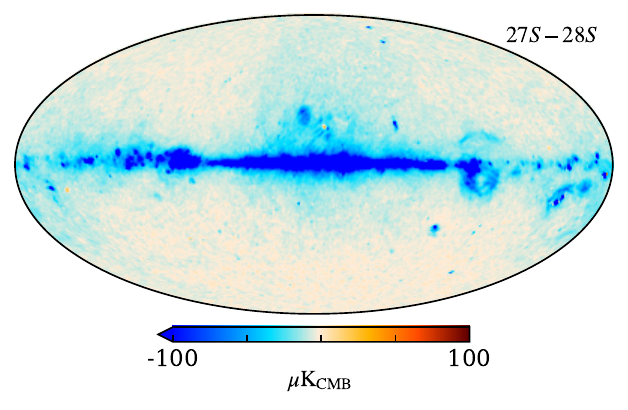}
\includegraphics[width=0.31\textwidth]{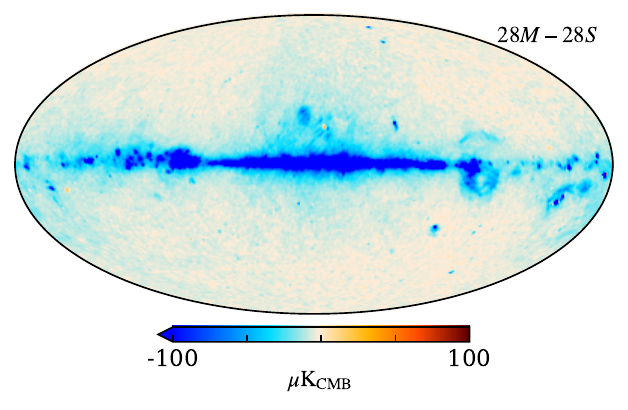}
\caption{All 6 possible difference maps between the simulated 30 GHz maps. 
	} \label{fig:sim_bp_diffs}
\end{figure*}

\begin{figure}[]
\includegraphics[width=\linewidth]{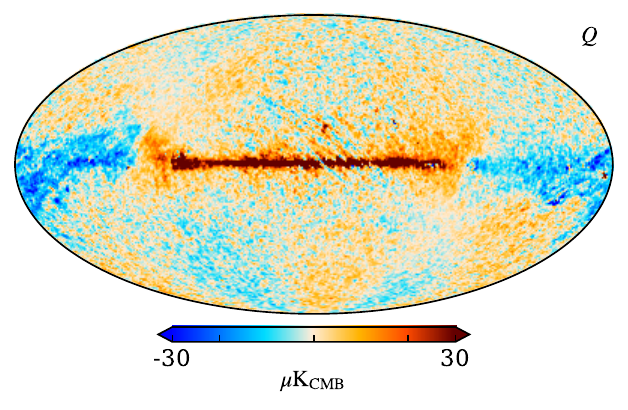}\\
\includegraphics[width=\linewidth]{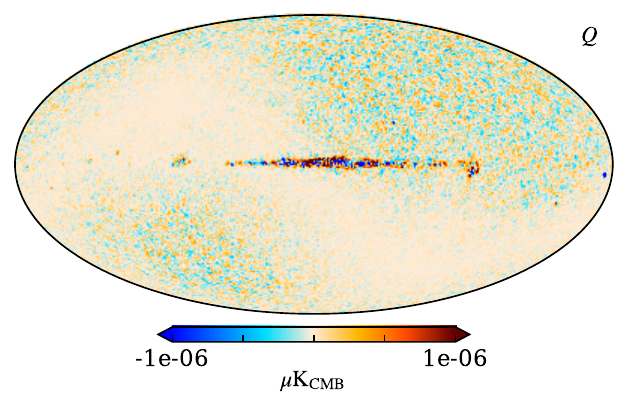}
\caption{Comparison of temperature-to-polarization leakage for a constant monopole term as processed through a standard co-adding mapmaker (top panel) and through the $N+2$ mapmaker (bottom panel). Note the different colour scales.}
\label{fig:leak}
\end{figure}

The original motivation for considering the $N+2$ mapmaking idea was twofold: firstly, the separation of the temperature maps for each detector improves our ability to perform component separation in the case of differing bandpasses, in particular for line emission components, and secondly that it helps mitigate temperature-to-polarization leakage caused by bandpass differences. To illustrate the first effect, Fig.~\ref{fig:sim_bp_diffs} shows pairwise temperature difference maps between the simulated temperature maps discussed above. We see that, for each case, the difference between channels $A$ and $B$ clearly exhibits the morphology of the Galactic plane, with an overall uniform color range; red if the effective central frequency for channel $A$ is lower than for channel $B$, and blue otherwise. This sign depends on the slope of the spectral energy density of the dominant foreground component in the Galaxy, which at 30\,GHz is synchrotron emission. This is stronger at lower frequencies, and thus a lower central frequency implies a stronger signal. These bandpass differences are precisely the effects that we want to capture with separate temperature maps.

To illustrate the second advantage, namely the capability of mitigating temperature-to-polarization leakage, we create yet another simulation of the same type as above (including polarization angle randomization), but this time we inject a bright artificial offset into one of the four 30\,GHz detectors, with an amplitude of 1\,mK. The signal is unpolarized, as it is added uniformly to each sample for the given detector, and therefore represents a detector-specific temperature excess. As a result, it will introduce temperature-to-polarization leakage, similar to for instance traditional bandpass, beam or gain mismatches. The top panel of Fig.~\ref{fig:leak} shows the Stokes $Q$ leakage map induced by this contribution when applying the traditional co-addition mapmaker, evaluated by subtracting the frequency maps derived with and without the spurious terms. For comparison, the bottom panel shows the same map when applying the $N+2$ mapmaker. The leakage amplitude in the second case is more than six orders of magnitude fainter. 

\section{Conclusions and Future Plans}
\label{sec:conclusions}

We have introduced $N+2$ mapmaking as a novel method producing temperature and polarization maps from multi-detector CMB timestreams. The motivation behind this method is two-fold; it allows the user to produce single-detector temperature maps from polarized TOD, which are useful for component separation purposes. Secondly, it helps mitigating various temperature-to-polarization leakage effects, for instance bandpass mismatch. Algorithmically speaking, this method is very closely related to the `spurious mapmaking' approach introduced by the \WMAP\ team, but rather than solving for $N-1$ residual maps, we solve directly for $N$ physically meaningful temperature maps. Indeed, the $N+2$ mapmaking algorithm was originally developed in preparation for a future Bayesian analysis of the \Planck\ HFI data, in which both leakage and CO line emimssion are important effects. This work is organized within the OpenHFI initiative, which is a sub-group of the Cosmoglobe collaboration. The next step in this process is to integrate the $N+2$ concept into an iterative Conjugate Gradient solver that simultaneously accounts for the HFI bolometer transfer function as demonstrated by \cite{artem}. 

In the current paper, we apply the $N+2$ formalism to the \Planck\ LFI 30\,GHz data within the Bayesian \commanderthree\ framework, at first attempting to produce single-detector temperature maps jointly with coadded polarization maps. Unfortunately, we find that poor polarization angle coverage of the \Planck\ scanning strategy prohibits a robust separation of temperature and polarization signals. In future work, it is instead worth considering producing horn maps, as for instance done in \Planck\ PR4 \citep{npipe}. However, in that work the horn maps were produced as a post-processing step that was separate from the main analysis, whereas with the new $N+2$ mapmaker the horn maps can be derived jointly with the co-added polarization maps. This is particularly useful in an iterative end-to-end Bayesian framework like \Cosmoglobe, in which close interaction between the calibration, mapmaking and component separation steps is essential.    

When applying the $N+2$ algorithm to a simulated LFI-like data set with randomized polarization angles, corresponding to adding a fast rotating half-wave plate to the instrument, we find that the $N+2$ algorithm produces maps that are consistent with expectations. In this case, we also find that the algorithm is capable of producing single-detector temperature maps with minimal temperature-to-polarization leakage. Incidentally, based on the same test, we also show that the introduction of a spinning half-wave plate does not by itself allow the production of clean polarization maps from multi-detector observations, which is of interest for future experiments that rely on a spinning half-wave plate, such as LiteBIRD \citep{litebird2022}. For these, the $N+2$ mapmaking algorithm offers the possibility of making multi-detector maps minimizing temperature-to-polarization leakage as an alternative to the standard single-detector maps. This could be useful for mitigating correlated but low signal-to-noise ratio systematic effects that would benefit from detector coaddition. 

\begin{acknowledgements}
 The authors would like to thank Dr. Sigurd Næss for useful conversations that he didn't think were sufficient to merit co-authorship.
 The current work has received funding from the European Union’s
 Horizon 2020 research and innovation programme under grant agreement
 numbers 819478 (ERC; \textsc{Cosmoglobe}), 101165647 (ERC;
 \textsc{Origins}), and 101141621 (ERC; \textsc{Commander}). This
 article reflects the views of the authors only. The funding body is
 not responsible for any use that may be made of the information
 contained therein. This research is also funded by the Research
 Council of Norway under grant agreement number 344934 (YRT;
 \textsc{CosmoglobeHD}). Some of the results in this paper have been
 derived using the HEALPix \citep{healpix} package.  We acknowledge
 the use of the Legacy Archive for Microwave Background Data Analysis
 (LAMBDA), part of the High Energy Astrophysics Science Archive Center
 (HEASARC). HEASARC/LAMBDA is a service of the Astrophysics Science
 Division at the NASA Goddard Space Flight Center.
 This paper and related research have been conducted during and with the support of the Italian national inter-university PhD programme in Space Science and Technology.
Work on this article was produced while attending the PhD program in PhD in Space Science and Technology at the University of Trento, Cycle XXXIX, with the support of a scholarship financed by the Ministerial Decree no. 118 of 2nd March 2023, based on the NRRP - funded by the European Union - NextGenerationEU - Mission 4 "Education and Research", Component 1 "Enhancement of the offer of educational services: from nurseries to universities” - Investment 4.1 “Extension of the number of research doctorates and innovative doctorates for public administration and cultural heritage” - CUP E66E23000110001.
\end{acknowledgements}

%-------------------------------------------------------------
%                                       Table with references 
%-------------------------------------------------------------
%

\bibliographystyle{aa}
\bibliography{references, CG_bibliography}
\end{document}